## Error-Rate Performance Analysis of Wireless Sensor Networks over Fading Channels

H. MEHDI++, S. SOOMRO*, W. R. KHAN, A. G. MEMON**, A. HAFEEZ*

Institute of Business and Technology, Karachi, Pakistan

**Abstract**: In this paper, we analyze the bit-error-rate (BER) performance of wireless sensor networks. A wireless sensor node with a single transmitter antenna and multiple receiver antennas is considered here. We consider $M$ ($M \geq 1$) receiver antennas to mitigate channel fading effects. BER performance is analyzed in the presence of a co-channel interference source. Wireless channel is assumed to follow independently Nakagami fading for the wireless sensor network and Rayleigh fading for the interference signal. Based on the derived analytical expressions, the effects of different fading channel and co-channel interference parameters on the BER performance of wireless sensor networks are assessed.

## I. INTRODUCTION

Wireless sensor networks are a quickly growing area for research and commercial progress. A wireless sensor network (WSN) is a collection of various data nodes deployed in a network. WSN is useful for military, environmental, and scientific applications to name a few. It is commonly composed of various wireless nodes that are distributed in a certain fashion to cover the area to be monitored (Constantin Volosencu and Daniel-Ioan Curiac, 2013). Instead of forwarding the raw data to other nodes in the network, sensor nodes use their processing abilities to carry out simple data processing and transmit only the required data. Therefore, WSN can be used for a wide range of applications (Yu, YushengJi, Liand Baohua and Zhao, 2012) and (Yuhong, and Wei Wayne LI, 2012). Due to wide spread use of the wireless sensor systems and due to limited frequency channels available the effects of co-channel interference is an important factor when designing systems based on wireless sensors (Abbosh and Thiel, 2005). Since various wireless sources may share spectrum with existing wireless sensor nodes, co-channel interference can be a limiting factor.To maintain the performance requirements in WSN, it is important that the effects of CCI be incorporated into the analysis of WSN. In this paper, our aim is to analyze the effects of CCI on the BER performance of WSN when there is a co-channel interference source within the vicinity of a sensor node. We assume interference mitigation techniques have not been incorporated in the system. In (Abbosh and Thiel, 2005), authors have discussed multiple-input multiple-output (MIMO) WSN systems with interference over a Rician channel. In (Kumar *et al.* 2011) and (Bao-Qiang Kanand, Jian-Huan 2012), authors have focused their work on the resource and interference management schemes. In this paper, our aim is to provide theoretical link in BER performance analysis of WSN in the presence of CCI over a Nakagami/Rayleigh fading channel. BER analysis is a well-known tool often used in the literature to analyze and study the performance of different types of systems under different channel conditions (Arunabha, *et. al.*, 2011).

In this paper, our objective is to analyze the wireless sensor networks with multiple antennas at the receiver side. The Nakagami distribution is used here to model the wireless channel for the sensor nodes. Space diversity with maximal-ratio-combining (MRC) technique is incorporated at the receiver side to combat fading conditions. We consider Nakagami distribution due to its ability to model various fading scenarios as well as a mathematically tractable form. We assume Rayleigh fading channel for the interferer signal. Rayleigh fading is applicable when there is no dominant propagation path along the line-of-sight between the transmitter and receiver. Our BER expressions are valid for arbitrary number of diversity branches, interference power and fading parameters.

## 2   MATERIALS AND METHODS

We consider a case where two wireless sensor nodes are in communication with each other. In this paper, we consider such a case in order to keep our analysis simple. In **(Fig. 1),** system layout of a two node communication scenario is shown, with one node is acting as a source and other node is acting as a receiver node. The receiver-node has $M$ ($M \geq 1$) receiver

++ Corresponding author E-mail: haiderpk0@gmail.com,
*Sindh Madressatul Islam University, Karachi
**Institute of Mathematics and Computer Science, University of Sindh

antennas to incorporate space diversity. We consider space diversity in order to combat fading conditions. The source-node employs single transmitter antenna. There is a single co-channel interferer in the system. This interferer may possibly be a wireless node acting as an independent co-channel interference source. Here, we consider binary-phase-shift-keying (BPSK) as the modulation technique. We consider BPSK in order to keep our mathematical analysis tractable. We assume a flat fading channel. We consider a Nakagami/Rayleigh fading channel in our theoretical analysis. The Nakagami distribution is expressed as (Arunabha , *et al.* 2011).

$$f_X(x) = \frac{2}{\Gamma(m_x)} \left(\frac{m_x}{\Omega}\right)^{m_x} x^{2m_x - 1} e^{-\frac{m_x x^2}{\Omega}}, \quad x > 0 \quad (1)$$

Where $E(X^2) = \Omega$, E(.) denotes the expectation operation and $\Gamma(.)$ is a gamma function (Gradshteyn, and Ryzhik, 2007). Nakagami fading parameter $m_x \in$ [0.5, ∞) in (1) controls the severity of fading. Less severe fading conditions are modeled by using larger values of the fading parameter $m_x$. The signal distribution for a Rayleigh fading channel is obtained for $m_x = 1$. The Nakagami fading parameter with $0.5 < m_x < 1$ models fading conditions more severe than that of Rayleigh fading. When $m_x > 1$, the model yields less severe fading conditions. The Nakagami distribution, therefore, provides a general model for a wireless system design and analysis. The Rayleigh distribution is expressed as (Arunabha , *et al.*, 2011).

$$f_X(x) = \frac{2}{\Omega} x e^{-\frac{x^2}{\Omega}}, \quad x > 0 \quad (2)$$

In the study of wireless signal propagation, wireless channel fading is often modeled by the Rayleigh distribution when there is no dominant propagation path along the line-of-sight between the source and the receiver. Here, we consider interference signal over a Rayleigh fading channel.

The received signals in each spatial diversity branch of the receiver after down-conversion, co-phasing and demodulation are combined by employing the diversity combining scheme based on the maximal-ratio-combining (MRC) principle (Arunabha, *et al.*, 2011). We assume perfect channel information is available at the receiver side. Thus, in each diversity branch of the receiver, the signal is weighted with respect to its respective channel gain. The final output signal-to-interference power ratio (SIR) can be written as

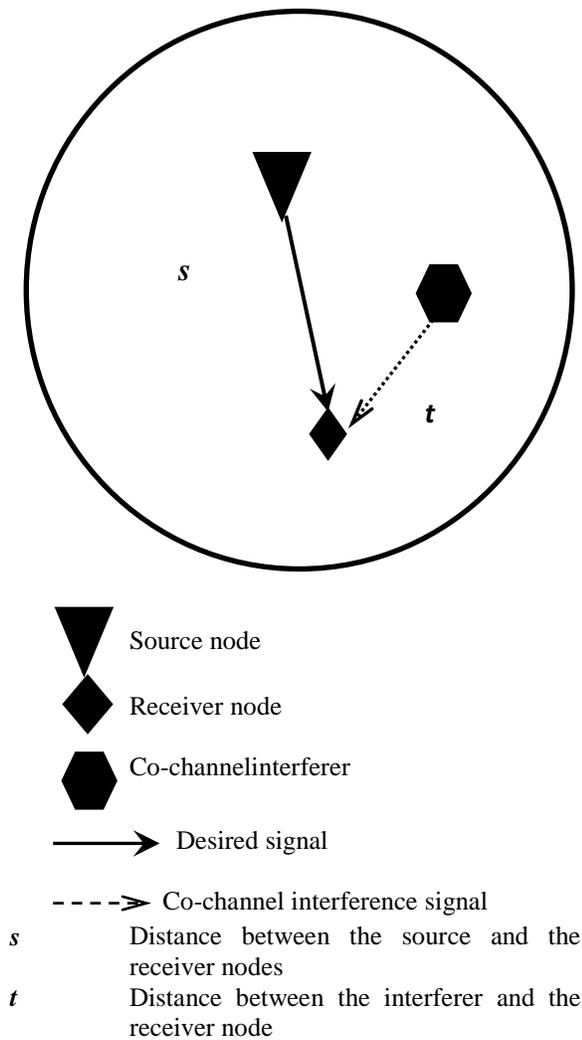

Source node

Receiver node

Co-channel interferer

⟶ Desired signal

----> Co-channel interference signal

*s*     Distance between the source and the receiver nodes

*t*     Distance between the interferer and the receiver node

**Fig. 1: System layout of a WSN with two node communication case in the presence of co-channel interference from a co-channel wireless node.**

$$\gamma = \frac{S}{I} = \frac{\sum_{v=1}^{M} h_v^2}{\frac{P_2}{P_1}\left(\frac{s}{t}\right)^n \alpha^2}, \quad (2)$$

Where $h_v$ is an independent Nakagami fading variable of the desired signal component from the transmitter and is received by the *v*-th diversity branch of the receiver. In (2), $P_1$ and $P_2$ are the transmitted powers of the source and the interferer nodes, respectively, *s* is the distance between the source and the receiver nodes, and *t* is the distance between the interferer and the receiver-nodes. The parameter *n* is the path-loss exponent (Arunabha *et al.*, 2011).

Random variable $\alpha$ is the independent Rayleigh fading channel variable of the co-channel interferer. Here, the effects of noise are ignored, because, we consider the interferer to be the only source of interference in the system. Now, by considering the formula,

$$f_\gamma(y) = \int_0^\infty z f_S(yz) f_I(z) dz,$$

and with the help of (Gradshteyn and Ryzhik, 2007) and (Yao and Sheikh, 1992), the probability density function (pdf) of SIR in (2), as a function of parameters $P_1$, $P_2$, $s$, $t$ and $n$ is given as:

$$f_\gamma(y) = \int_0^\infty z \underbrace{\frac{\left(\frac{m}{\sigma}\right)^{Mm}}{\Gamma(Mm)} y^{(Mm-1)} z^{(Mm-1)} e^{\left(-\frac{m}{\sigma}yz\right)}}_{f_S(yz)} \times \underbrace{\frac{1}{\frac{P_2}{P_1}\left(\frac{s}{t}\right)^n \rho} e^{\left(-\frac{1}{\frac{P_2}{P_1}\left(\frac{s}{t}\right)^n \rho}z\right)}}_{f_I(z)} dz,$$

$$f_\gamma(y) = \left(\frac{1}{\frac{P_2}{P_1}\left(\frac{s}{t}\right)^n \rho}\right) \left(\frac{m}{\sigma}\right)^{Mm} \times \frac{y^{(Mm-1)}}{\Gamma(Mm)} \int_0^\infty z^{Mm} \exp\left[-\left(\frac{1}{\frac{P_2}{P_1}\left(\frac{s}{t}\right)^n \rho} - \frac{m}{\sigma}y\right)z\right] dz,$$

$$f_\gamma(y) = \left(\frac{1}{\frac{P_2}{P_1}\left(\frac{s}{t}\right)^n \rho}\right) \left(\frac{m}{\sigma}\right)^{Mm} Mm \times y^{Mm-1} \left(\frac{1}{\frac{P_2}{P_1}\left(\frac{s}{t}\right)^n \rho} + \frac{m}{\sigma}y\right)^{-(Mm+1)},$$

Where $\Gamma(.)$ is a gamma function (Gradshteyn and Ryzhik, 2007), $\sigma = E(h_v^2)$ and $\rho = E(\alpha^2)$. In (3), *M* is the number of receiver antennas at the receiver node side. In what follows, based on (3), we will present the BER expression. The bit error probability is a well-known tool often used in the literature to analyze the performance of different types of communication systems under various channel conditions. The BER is given by (Simon, and Alouni, 2004)

$$P = \frac{1}{2\Gamma(0.5)} \int_0^\infty \Gamma(0.5, y) f_\gamma(y) dy, \quad (4)$$

Where $\Gamma(.,.)$ denotes the complementary incomplete gamma function (Gradshteyn and Ryzhik, 2007). In (4), $f_\gamma(.)$ is the pdf given in (3). The expressions (3) and (4) are valid for arbitrary number of diversity branches, interference power and fading parameters. Furthermore, the level of integration in (4) is independent of the number of diversity branches, i.e., only a single level of integration is needed. The expressions (3) and (4) are functions of distance between the source and the receiver nodes, i.e., *s*, the distance between the co-channel interferer and the receiver node, i.e., *t*, and the path-loss exponent, i.e., *n*. In the subsequent section, we will analyze the system performance in terms of these parameters.

## 3. NUMERICAL RESULTS AND DISCUSSION

In this section, we will present and discuss numerical results based on the expressions derived in Section II. In **(Fig. 2)**, BER performance is shown. We assume the path-loss exponent to be 3.5. We assume $P_1$ and $P_2$ to be 17 dBm and 10 dBm, respectively. We fix the number of diversity branches to *M* = 2. We consider the Nakagami fading parameter for the sensor network to be 3. We vary the distance between the receiver node and the co-channel interferer, i.e., *t* and, the distance between the receiver and the source nodes, i.e., *s*. From **(Fig. 2)**, we observe that by varying the distance between the receiver and the source nodes the BER performance varies. It is due the fact that as the distance between the receiver and the source nodes is increased the power received by the receiver is decreased due to path loss conditions. The received power at the receiver node from the source node has an inverse relation with the distance values, i.e., *s* or *t* (Arunabha, *at el.* 2011). Also, as the value of *t* is decreased, the received power of the co-channel interferer is increased at the receiver, and thus, overall system SIR is decreased. And thus we observe degradation in the BER performance due to increase in interference power at the receiver node.

In **(Fig. 3),** the BER performance is shown with various combination of number of diversity receiver antennas, i.e., *M*. We consider the Nakagami fading parameter for the sensor network to be 2. We assume the path-loss exponent to be 3.0. We assume $P_1$ and $P_2$ to be 15 dBm and 6 dBm, respectively. We fix the distance between the receiver node and co-channel interferer to be 90 meters. From the figure, we observe improvement in the BER performance as the number of receiver antennas increases, due to the improvement in SIR conditions. In **(Fig. 4),** we compare the BER performance under varying conditions of co-channel interference power. We assume system path-loss exponent to be *n* = 2.9. We consider the Nakagami fading parameter for the sensor network to be 4. We fix the number of receiver antennas to be *M* = 3. The distance between the receiver-node and co-channel interferer is assumed to be 80 meters. We assume $P_1$ to be 15 dBm. From the figure, we observe that as the value of interference power decreases BER improves. It is due to the reason that as the value of interference power decreases, the SIR condition is improved and thus overall system BER performance is improved.

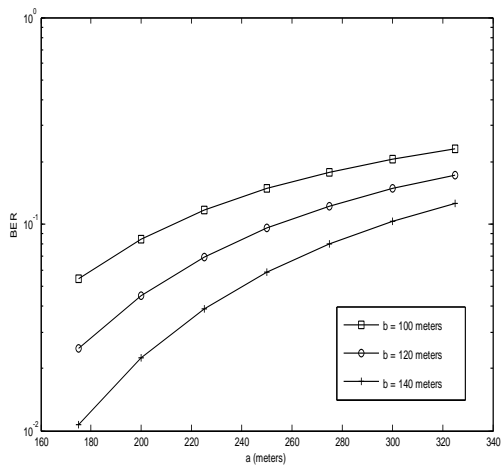

**Fig. 2: BER performance of wireless sensor networksbyvarying the distance betweenco-channel interfererand the receiver node.**

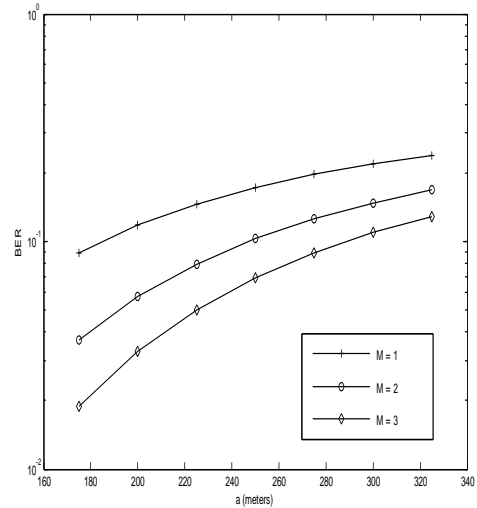

**Fig. 3: BER performance of wireless sensor networksunder various diversity conditions at the receiver node.**

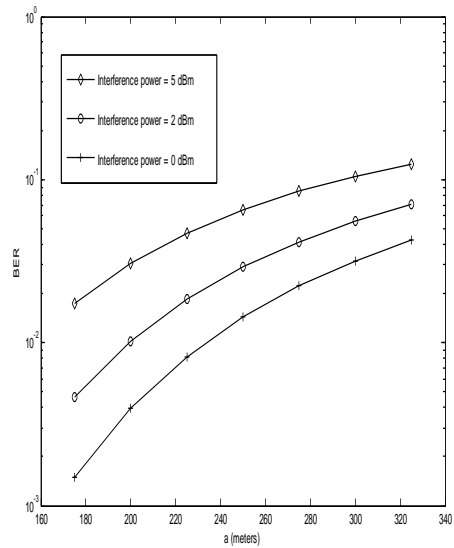

**Fig. 4: BER performance of wireless sensor networkswith various interference power levels.**

**4.** **CONCLUSION**

The BER performance of a wireless sensor network has been studied in the presence of co-channel interference. We study the BER performance in the presence of a wireless node acting as a co-channel interference. From our numerical results, we observe degradation in the error-rate performance of the system due to the presence of co-channel interference. This degradation worsens as the receiver node moves farther from the source node and nearer to the interference source. We further observe improvement in the BER performance when the number of the diversity branches is increased.

**REFERENCES:**

Arunabha G., J. Zhang, J. G. Andrews and R. Muhamed, 2011, Fundamentals of LTE, Boston, MA, Pearson Education, 2011.

Abbosh M., and D. Thiel, (2005) "Performance of MIMO-Based wireless sensor networks with co-channel interference," in Proc. IEEE Conf. On Intelligent Sensors, Sensor Networks and Information Processing, Melbourne, Australia, 115–119.

Bao-Qiang K. and F. Jian-Huan (2012), "Interference activity aware multi-path Routing protocol," EURASIP Journal on Wireless Communications and Networking Aug. 2012.
*http://jwcn.eurasipjournals.com/content/2012/1/267*

Constantin V. and C. Daniel-Ioan (2013) "Efficiency improvement in multi-sensor wireless network based estimation algorithms for distributed parameter systems with application at the heat transfer,"EURASIP Journal on Advances in Signal Processing, January 2013.
*http://asp.eurasipjournals.com/content/2013/1/4*

Gradshteyn I. S. and I. M. Ryzhik, (2007) Tables of Integrals, Series and Products, San Diego, CA, Academic, 7th ed., 2007.

Kumar S., A. Sharma and S. S. Raghuvanshi, (2011) "Energy efficient scheduling algorithm With interference reduction for wireless sensor networks, "in Proc. IEEE Conf. on Computational Intelligence and Communication Networks (CICN), Gwalior, India 328-332.

Ping R, (2010) "A nested cellular topology applicable to wireless sensor networks," in Proc. 12th IEEE International Conference on Communication Technology (ICCT), Nanjing, China, 689-692, 11-4 Nov. 2010.

Yao Y. D. and A. U. H. Sheikh, (1992) "Investigations into cochannel interference in microcellular mobile radio systems, "IEEE Trans. Veh. Technol., Vol. 41, 114-123, May 1992.

Simon, M. K. and M. S. Alouni, (2004) Digital Communication over Fading Channels, New York, NY, John Wiley and Sons Inc., 2nd ed., 2004.

Yu G., J. YushengJi, Z. Liand Baohua (2012) "Towards an optimal lifetime in heterogeneous Surveillance wireless sensor networks," EURASIP Journal on Wireless Communications and Networking, March 2012.
*http://jwcn.eurasipjournals.com/content/2012/1/74*

Yuhong Z. and L I. W. Wayne (2012) "Modeling and energy consumption evaluation of a stochastic wireless sensor network," EURASIP Journal on Wireless Communications and Networking.
*http://jwcn.eurasipjournals.com/content/2012/1/282*